\documentclass[twocolumn,showpacs,preprintnumbers,amsmath,amssymb]{revtex4}
\usepackage{graphicx}
\usepackage{dcolumn}
\usepackage{bm}
\usepackage{comment}
\usepackage{subfigure}
\usepackage{amsmath}
\usepackage{url}
\usepackage{hyperref}
\usepackage{breakurl}

\begin{document}

\title{Octupolar out-of-plane magnetic field  structure generation during collisionless magnetic reconnection in a stressed $X$-point collapse} 
\author{J. Graf von der Pahlen and D. Tsiklauri}
\affiliation{School of Physics and Astronomy, Queen Mary University of London, London, E1 4NS, United Kingdom}
\date{\today}
\begin{abstract}
The out-of-plane magnetic field, generated by fast magnetic reconnection, during collisionless, stressed $X$-point collapse, was studied with a kinetic, 2.5D, fully electromagnetic, relativistic particle-in-cell numerical code, using both closed (flux conserving) and open boundary conditions on a square grid. It was discovered that the well known quadrupolar structure in the out-of-plane magnetic field gains four additional regions of opposite magnetic polarity, emerging near the corners of the simulation box, moving towards the $X$-point. The emerging, outer, magnetic field structure has opposite polarity to the inner quadrupolar structure, leading to an overall octupolar structure. Using Ampere's law and integrating electron and ion currents, defined at grid cells, over the simulation domain, contributions to the out-of-plane magnetic field from electron and ion currents were determined. The emerging regions of opposite magnetic polarity were shown to be the result of ion currents. Magnetic octupolar structure is found to be a signature of $X$-point collapse, rather than tearing mode, and factors relating to potential discoveries in experimental scenarios or space-craft observations are discussed.  
\end{abstract}

\pacs{52.65.Rr;52.30.Cv;52.27.Ny;52.35.Vd;52.35.Py}

\maketitle

\pagestyle{plain}
\thispagestyle{plain}

Magnetic Hall reconnection, as first proposed by B. Sonnerup in 1979 \cite{sonnerup}, is a mode of reconnection relying on the decoupling of ions and electrons in a diffusion region and is of great interest in the study of magnetic reconnection as it presents an alternative to the the Petschek model, which relies on an anomalous resistivity \cite{Erkaev}. Even in setups suitable for Petschek reconnection, contributions of Hall effects need to be considered. A recent analytical result, corroborated by a numerical study, shows that the transition from Petschek to Hall reconnection occurs when the half-length of the current sheet reaches the ion inertial length \cite{arber}. This was shown as a direct consequence of a generalised scaling law, relating the reconnection rate to the distance between the $X$-point and the start of slow mode shocks.

An observational consequence of Hall reconnection is the associated quadrupolar out-of-plane magnetic field, induced by currents resulting from the decoupling i.e. the Hall currents, first demonstrated in a study by Teresawa in 1983 \cite{terasawa}. From 1994, the effect was further shown to occur in numerical Hybrid simulations \cite{mandt,karimabadquad,arznerquad} and later in a full Particle In Cell (PIC) numerical simulation in 2001 \cite{prittchet}. However, as shown in Ref. \cite{karimabadioct}, magnetic reconnection can lead to quadrupolar magnetic field structure generation, even without the Hall effect. By being an observational signature of magnetic reconnection, the quadrupolar field has thus been of great interest in recent  spacecraft missions, including Polar \cite{Mozer} in 2002 and Cluster \cite{Borg} in 2005, which both observed individual magnetic poles in the magnetotail of the Earth. Subsequently, a full quadrupolar pattern was observed in a multi-spacecraft Cluster mission in 2007 \cite{Eastwood}. The experimental evidence of the full quadrupolar structure was found at the MRX facility in 2004 \cite{mrx}. A more comprehensive account of the developments in Hall reconnection, as well as a review of relevant theory, can be found in Ref. \cite{Uzdensky}. Here, the discovery of a related effect, leading to an additional four regions of opposite magnetic polarity in the out-of-plane magnetic field is presented. The resulting overall magnetic field has an octupolar structure, which could be of similar observational significance and a potential avenue for further experimental investigation. We shall refer to the central quadrupolar magnetic field structure as quadrupolar components and to the additional regions of magnetic polarity as octupolar components.

The reconnection setup used in this study is that of $X$-point collapse, first introduced by Dungey in 1953 \cite{Dungey53}, as one of the earliest analysis of magnetic reconnection, pre-dating the tearing-mode. While this setup is distinctly different from  the well-studied tearing-mode instability, after a Harris type current sheet is disrupted by the tearing instability and magnetic islands and $X$-points start to form, there are few distinguishable differences between the two approaches. In both cases a stage is reached where $X$-point symmetry is broken, which means that there is no restoring force and the $X$-point collapses, resulting in fast reconnection (see Ref. \cite{priest}, chapter 7.1). Moreover, even the respective causes of the reconnection electric field, as calculated using the generalized Ohm's law, are the same, namely the off-diagonal terms of the divergence of the electron pressure tensor. This was shown in Ref. \cite{tsiklauri2008} for the case of $X$-point collapse and in Ref. \cite{prittchet} for the tearing-mode instability. The relation of these 2D simulation results to 3D reconnection was studied in Ref. \cite{karimabadi}, where it was found that 2D magnetic islands manifest themselves as the flux tubes. Particle acceleration, due to 3D effects in resistive MHD reconnection, was studied in Ref. \cite{browning}.

\begin{figure}[htbp]
    \includegraphics[width=0.99\linewidth]{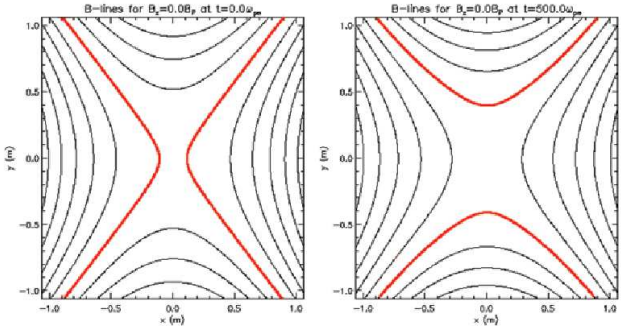}
     \caption{ \label{Fig:fieldcollapse}Magnetic field line configuration at the start of the simulation (left) and at the end of the simulation (right), for closed boundary condition at a system size of $~4c/\omega_{pi}$. Note that, while field lines reconnect and shift, as indicated by the change in the thick red line from the left to the right panel, all field lines remain at the same location on the boundary, as is dictated by the flux-conserving boundary conditions.} 
   \end{figure}

Previous works on collisionless $X$-point collapse can be found in \cite{tsiklauri2008,Tsiklauri2007,gvdp}. The setup of the initial in plane field is a standard $X$-point configuration, given by

    \begin{equation}
    B_x = \frac{B_0}{L} y,
    \;\;\;    B_y = \frac{B_0}{L} \alpha^2 x, \;\;\; 
    \label{eqn:Bxy}
    \end{equation}
where $B_0$ is magnetic field intensity at the distance $L$ from the $X$-point for $\alpha = 1.0$, $L$ is the global external length-scale of reconnection, and $\alpha$ is the stress parameter (see e.g. chapter 2.1 in Ref. \cite{priest}). The left panel of Fig.~\ref{Fig:fieldcollapse} shows this initial setup for the in-plane magnetic field.  In addition a uniform current is imposed at time $t=0$ in the $z$-direction, corresponding to the curl of the magnetic field, such that Ampere's law is satisfied

  \begin{equation}
    j_z = \frac{B_0}{\mu_0 L} (\alpha^2 - 1).
   \end{equation}

In this scenario the initial stress in the field leads to a $\bf{J}\times\bf{B}$ force that pushes the field lines horizontally inwards. This serves to increase the initial imbalance, which in turn increases the inwards force and the field collapses. Due to the frozen-in condition, this leads to a build up of plasma near the $X$-point and eventually to the formation of a diffusion region, accompanied by a current sheet. The final configuration of the field lines can be seen in the right panel of Fig.~\ref{Fig:fieldcollapse}. The term in the generalized Ohm's law, which results in the breaking of the frozen in condition, was shown to be the off-diagonal terms of the electron pressure tensor divergence, due to electron meandering motion (see Ref. \cite{tsiklauri2008}). This makes the reconnection process fast compared to the resistive MHD, which is too inefficient (see Ref. \cite{priest}, chapter 7.1.1).

The simulation code used here is a 2.5D relativistic and fully electromagnetic Particle In Cell (PIC) code, developed by the EPOCH collaboration, based on the original PSC code by Hartmut Ruhl \cite{ruhl2006} (minor modifications were made in this study to allow for flux conserving boundary conditions). The number density of both electrons and ions in the simulation domain, $n_e$ and $n_p$, was set to $10^{16}m^{-2}$, while the temperature for both electrons and protons, $T_e$ and $T_p$, was set to $6.0\times10^7$K, matching conditions of flaring in the solar corona. The proton mass was set to 100 times the electron mass, i.e. $m_p = 100m_e$, to speed up the code. 
 
Two types of boundary conditions were used: closed boundary conditions, which conserve flux at the boundary and reflect particles, and open boundary conditions, which allow in and out flow of magnetic flux and remove outgoing particles, reaching the boundary, from the system. In the closed case, zero-gradient boundary conditions are imposed both on the electric and magnetic fields in  $x$- and $y$-directions and the tangential component of electric field was forced to zero, while the normal component of the magnetic field was kept constant. This anchors magnetic field lines on the boundary (see Fig.~\ref{Fig:fieldcollapse}), thus inhibiting loss (or gain) of magnetic flux from the simulation domain. This is applicable to magnetic fields which are anchored in the solar photosphere by the frozen-in condition and form an X-point higher up in the corona, which serves as a simplest model of a solar coronal active region. Open boundary conditions on the other hand mimic a large open system and thus more representative of scenarios such as reconnection in the magnetotail of the Earth. As described in \cite{ruhl2006}, chapter 4.2.3, waves are transmitted out of the boundary by decomposing the governing equations into their forwards and backwards propagating wave components and setting these to be a constant across the boundary. Here, parallel components of the magnetic field are held fixed at the boundary, effectively letting the system evolve as if embedded in a magnetic field structure, providing a field line inflow/outflow such that the parallel field at the boundary remains constant (tests with zero-gradient boundary conditions on magnetic fields, as studied in \cite{zerograd}, also result in similar generated octupolar out-of-plane magnetic field structure).   



For both boundary conditions, different system sizes were used for the square simulation domain, ranging from $~4c/\omega_{pi}$ to $~16c/\omega_{pi}$, in order to investigate how the reconnection and associated effects vary with spatial scale. The effective grid sizes used ranged from $2L=2.14$ m to $2L=8.56$ m. Simulation grid cells were set to the Debye length, thus leading to grids ranging from 400x400 to 1600x1600 grid cells. The code was set to use 200 particles per species per cell, which was shown to be sufficient for accurate representations of the EM field dynamics in convergence tests. The value of $B_0$ for the different runs was adjusted such that the Alfv\'en speed at the $y$-boundary was fixed as $v_a = B_b/\sqrt{\mu_0\rho} = 0.1c$, where $B_b$ represents the strength of the magnetic field at $(x_{max},0)$. For meaningful comparison between the runs, the $x$-axes of plots showing time dynamics use Alfv\'en time, $t_a=L/v_a$.

\begin{figure}[htbp]
    \includegraphics[width=0.99\linewidth]{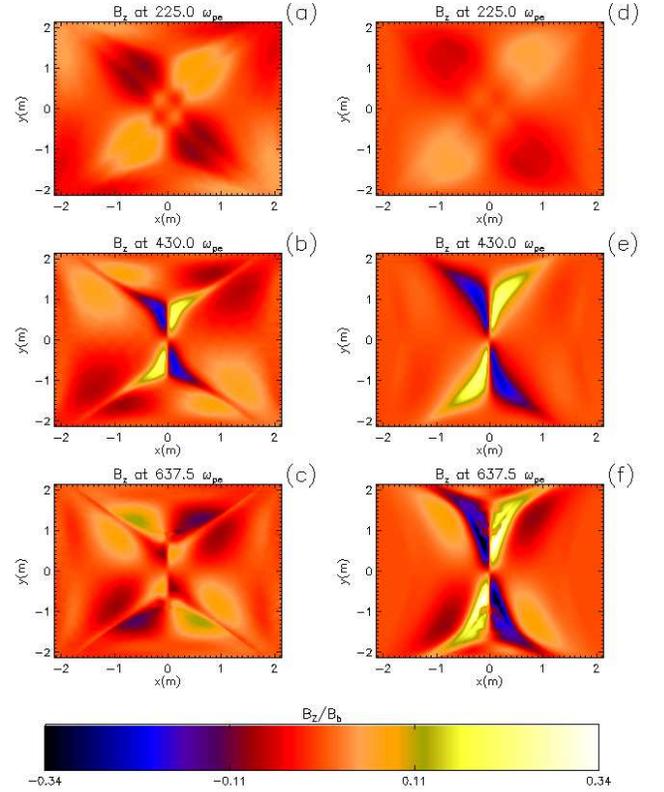}
     \caption{Vertically from (a) to (c), the out-of-plane magnetic field at 3 consecutive times, for closed boundary conditions for a domain size of $~8c/\omega_{pi}$. Note that at the onset of reconnection, in panel (a), an initial octupolar field pattern occurs, similar to the one shown in Fig. 5 in Ref. \cite{karimabadioct}, while later in the simulation a more substantial octupolar field emerges, having opposite polarity to the initial one.  Again vertically, panels (d) to (f) similarly show the evolution of the out-of-plane magnetic field for open boundary conditions.}  
     \label{Fig:octosnaps}
   \end{figure}

\begin{figure}[htbp]
    \includegraphics[width=0.99\linewidth]{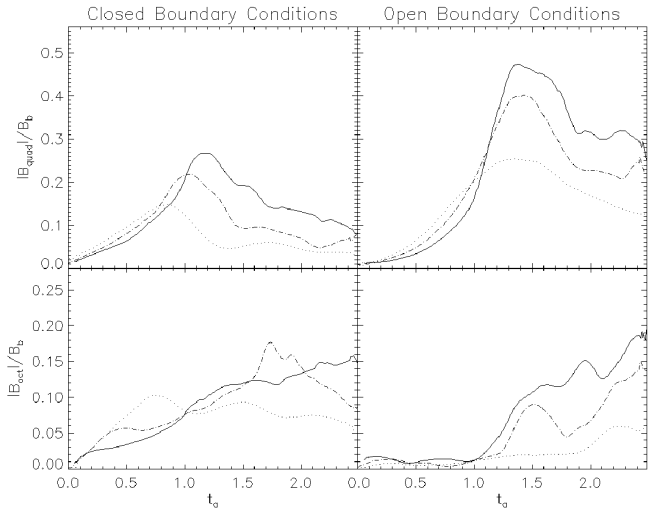}
  \caption{Showing time dynamics of octupolar and quadrupolar magnetic field component maxima for closed and open boundary conditions. Dotted lines correspond to system lengths of $~4c/\omega_{pi}$, dashed lines to system sizes of $~8c/\omega_{pi}$ and solid lines to system sizes of $~16c/\omega_{pi}$.}
  \label{Fig:bzquadoct}
\end{figure}

For both open and closed boundary conditions and for all system sizes, the reconnection rate and the associated quadrupolar magnetic field peak and subsequently decline within 2.5$t_{\alpha}$ (or 500$\omega_{pe}^{-1}$, 1000$\omega_{pe}^{-1}$ and 2000$\omega_{pe}^{-1}$ for the three domain sizes.) This is shown for the $~8c/\omega_{pi}$ system size cases in Fig.~\ref{Fig:octosnaps}. As shown, after about 400$\omega_{pe}^{-1}$ in the closed case and 600$\omega_{pe}^{-1}$ in the open case, additional regions of magnetic polarity emerge near the corners of the domain, each with opposite polarity of the respective region of the quarupolar field at the centre of the domain. In order to determine the strengths of the magnetic field components making up the quadrupole (quadrupolar components) and the additional ones (octupolar components), the bottom left quarter of the simulation domain was isolated and the maximum value of the out-of-plane magnetic field, representative of the quadrupolar field, and the minimum value, representative of the octupolar field components, were plotted (see Fig.~\ref{Fig:bzquadoct}). As shown, for both boundary cases, the octupolar field components reach a peak in field strength only after the peak in the quadrupolar field is reached. Also, it is shown that increasing the domain size leads to an increase in the strength of octupolar field components, indicating that this effect could occur in a wide open system. Excluding the small scale open boundary case, where no significant development was observed, peak field strengths of octupolar components are shown to range from $0.1B_b$ to $0.2B_b$, representing a significant fraction of the quadrupolar field strength in both boundary cases. 

It is to be noted that, at the beginning of the simulation, a different type of octupolar magnetic field structure emerges, as shown in panel (a) of Fig.~\ref{Fig:octosnaps}, where regions of opposite field polarity briefly appear within the quadrupolar structure, at a significantly smaller magnetic field strength. The same effect was demonstrated in Ref. \cite{karimabadioct} using a hybrid simulation of a tearing instability and was shown to be the result of a competition between initial differential ion flows. It is thus confirmed that this effect occurs both in $X$-point collapse and tearing-mode reconnection setups.

To determine the cause of the additional magnetic poles in terms of the currents in the simulation, Ampere's law was taken in component form, such that
    \begin{equation} 
    dB_z=\mu_0j_{x,ion}dy+\mu_0j_{x,electron}dy+\frac{1}{c^2}\frac{\partial E_x}{\partial t}dy 
        \label{eqn:ampere1}
    \end{equation}
    and
        \begin{equation} 
    dB_z=-\mu_0j_{y,ion}dx-\mu_0j_{y,electron}dx-\frac{1}{c^2}\frac{\partial E_y}{\partial t}dx.
    \label{eqn:ampere2}
    \end{equation}
    
\begin{figure}[htbp]
 \includegraphics[width=0.99\linewidth]{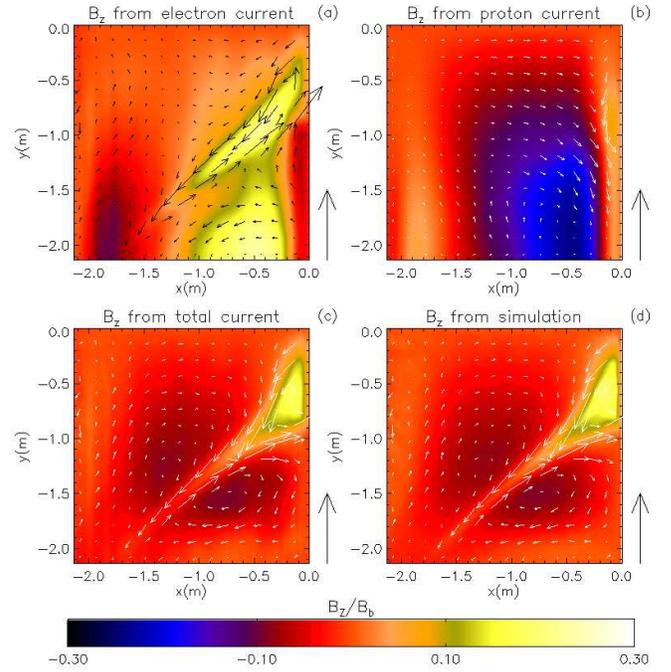}
 \caption{For closed boundary conditions for a domain size of $~8c/\omega_{pi}$ around $430\omega_{pe}^{-1}$ ($1.08t_a$), when octupolar components first emerge, panels (a) and (b) show the out-of-plane magnetic field over the bottom left quadrant of the domain, calculated from electron and ion currents respectively, based on Ampere's law. Panel (c) shows the total of these two contributions as well as the contribution from the displacement current (small). Panel panel (d) shows the full out-of-plane magnetic field obtained directly from the simulation, effectively representing the bottom right quadrant of panel (b) in Figure~\ref{Fig:octosnaps}. Arrows on panels indicate direction and intensity of the current density at respective grid cells. Arrows beside panels represent the current density for charged particles moving at the Alven speed ($v_a = 0.1c$) with a particle density of $n_e$, i.e. the initial charge density in the simulation domain.} 
 \label{Fig:bzj}
\end{figure}

This way an integration could be carried out, allowing for $B_z$ to be derived from the individual currents, allowing contributions from different currents to be established. The individual currents for electrons and ions are calculated by EPOCH on each grid cell and the displacement current was obtained by taking a five-point stencil using electric field values at the same cell, separated over four time steps. As a starting point for the integration the new $B_z$ was set to zero at the centre of the grid, i.e. $B_z(0,0)=0.0$, as is consistent with it being at the centre of the domain. Thus, it was possible to individually integrate over the simulation grid, using the three different currents, to obtain their individual contributions to $B_z$. E.g. using Eq. (\ref{eqn:ampere1}) one obtains

    \begin{equation} 
B_{z,ion}(0,L_y)=\int_0^{L_y} \! j_{x,ion}(0,y) \, \mathrm{d}y 
    \label{eqn:ampere3}
    \end{equation}
    
followed by (\ref{eqn:ampere2}) to get

    \begin{equation} 
B_{z,ion}(L_x,L_y)=-\int_0^{L_x} \! j_{y,ion}(x,L_y) \, \mathrm{d}x, 
    \label{eqn:ampere4}
    \end{equation}
    
 where $(L_x,L_y)$ represents an arbitrary point on the $B_z$ grid. Carrying out the same integration for all $L_x$ and $L_y$ on the grid also for the electron and the displacement currents, plots shown in Fig.~\ref{Fig:bzj} are obtained. Only the lower left quarter of the simulation domain is shown due to considerations of symmetry. As expected, based on Hall dynamics, the contribution to the quadrupolar components is provided entirely by the electron currents. As explained in Ref. \cite{Uzdensky} they are the result of coupled electrons moving to and from the $X$-point in order to conserve charge neutrality as field lines deform during reconnection. In addition to this, we show that the contribution to the octupolar components is provided by the ion currents. A further observation is that, after being expelled from the $X$-point, there is an additional flow of electrons towards the horizontal centre of the domain. This is likely also due to electrons moving along the field lines to restore charge neutrality. This current makes up the additional positive contribution to the out-of-plane magnetic field at the bottom of panel (a) in Fig.~\ref{Fig:bzj} and it is this contribution which leads to the separation of the octupolar components from the simulation boundary (see panel (b) to (c) in Figure~\ref{Fig:octosnaps} ). The contribution of the displacement currents was small in comparison and therefore a dedicated plot is omitted. By comparing panels (c) and (d) of Fig.~\ref{Fig:bzj}, it can be seen that the calculated field corresponds well with the one obtained directly from the simulation.

In conclusion: in addition to the quadrupolar magnetic field, known to emerge in magnetic reconnection, four additional regions of opposite magnetic polarity in the out-of-plane magnetic field were observed in a PIC simulation, using a stressed $X$-point collapse, for both closed and open boundary conditions. As shown in Fig.~\ref{Fig:octosnaps}, for each central region of magnetic polarity, an additional region of opposite magnetic polarity emerges. The resulting octupolar structure appeared most prominently in late in the simulation, i.e. after the peak in the quadrupolar field strength was reached. This is consistent with plots obtained from taking the maximum and minimum out-of-plane magnetic field values in the lower left quarter of the domain, representative of the quadrupolar and octupolar field components respectively (see Fig.~\ref{Fig:bzquadoct}). For both boundary conditions, the quadrupolar field components consistently peak and diminish within 2.5$t_a$, while on the other hand the octupolar field components remained substantial. Thus, there exists a point when both magnetic fields are approximately equal. An ideal opportunity for observation would thus be at later times, when quadrupolar components subside, while octupolar components persist. 

By breaking Ampere's law into components given on the simulation grid and integrating to obtain $B_z$ it was possible to determine contributions to $B_z$ from electron and proton currents separately (see Fig.~\ref{Fig:bzj}). It was shown that the inner quadrupolar structure is linked to the electron motion, as is consistent with Hall dynamics, while the octupolar components are linked to the ion motion. This can be explained by the fact that, while electrons are coupled to field lines shortly after reconnection, their vertical motion away from the $X$-point thus determined by the out flow speed of the field lines, ions travel a greater distance before recoupling to the field. As shown in panel (a) of Fig.~\ref{Fig:bzj}, electrons move along the field lines through the $X$-point and no significant motion perpendicular to the field lines occurs (note that current direction is opposite to electron flow direction). Panel (b) on the other hand shows ions moving across the field lines towards and away from the $X$-point, thus resulting in the current making up the octupolar field components. As the reconnection rate, and thus the velocity of out-flowing field lines decreases, a point is reached where contributions from the ion current become dominant. 

Since the magnetic quadrupole serves as a marker of Hall reconnection it would be reasonable to investigate if the magnetic octupolar structure presented here could also be observed in a laboratory or in space-craft missions. It is to be stressed that, as shown in Fig.~\ref{Fig:bzquadoct}, the magnetic components making up the octupolar field become more significant after the inner quadrupolar field has peaked. Initial testing with greater ion to electron mass ratios further indicate that the latter is the case and will be reported elsewhere. We also stress that open boundary conditions are applicable to geomagnetic tail reconnection. In fact, while this was studied with focus on tearing mode, Dungey's first model was based on $X$-point collapse. Thus we believe that observing octupolar structure in the geomagnetic tail could be a distinguishing factor between tearing mode and $X$-point collapse reconnection models.      


\acknowledgements
  Authors acknowledge use of Particle-In-Cell code EPOCH and support by development team (http://ccpforge.cse.rl.ac.uk/gf/project/epoch/). Computational facilities used are that of Astronomy Unit, Queen Mary University of London and STFC-funded UKMHD consortium at St. Andrews and Warwick Universities. JGVDP acknowledges support from STFC PhD studentship. DT is financially supported by STFC consolidated Grant ST/J001546/1, The Leverhulme Trust Research Project Grant RPG-311 and HEFCE-funded South East Physics Network (SEPNET).


\end{document}